# Atomic-scale visualization of surface-assisted orbital order

Howon Kim[1]†, Yasuo Yoshida[1]*, Chi-Cheng Lee[2,3]‡, Tay-Rong Chang[4], Horng-Tay Jeng[4,5], Hsin Lin[2,3], Yoshinori Haga[6], Zachary Fisk[6,7], Yukio Hasegawa[1]

**Affiliations:**

[1]Institute for Solid State Physics, University of Tokyo, Kashiwa 277-8581, Japan.

[2]Centre for Advanced 2D Materials and Graphene Research Centre, National University of Singapore, Singapore 117546, Singapore.

[3]Department of Physics, National University of Singapore, Singapore 117542, Singapore.

[4]Department of Physics, National Tsing Hua University, Hsinchu 30013, Taiwan.

[5]Institute of Physics, Academia Sinica, Taipei 11529, Taiwan.

[6]Advanced Science Research Center, Japan Atomic Energy Agency, Tokai 319-1195, Japan.
[7]Department of Physics and Astronomy, University of California, Irvine, California 92697, USA.

*Correspondence to: yyoshida@issp.u-tokyo.ac.jp

† Present address: Department of Physics, University of Hamburg, Jungiusstrasse 11, D-20355 Hamburg, Germany

‡ Present address: Institute for Solid State Physics, University of Tokyo, Kashiwa 277-8581, Japan.

**Abstract**: Orbital-related physics attracts growing interest in condensed matter research, but direct real-space access of the orbital degree of freedom is challenging. Here we report a first, real-space, imaging of a surface-assisted orbital ordered structure on a cobalt-terminated surface of the well-studied heavy fermion compound $CeCoIn_5$. Within small tip-sample distances, the cobalt atoms on a cleaved (001) surface take on dumbbell shapes alternatingly aligned in the [100] and [010] directions in scanning tunneling microscopy topographies. First-principles calculations reveal that this structure is a consequence of the staggered $d_{xz}$-$d_{yz}$ orbital order triggered by enhanced on-site Coulomb interaction at the surface. This so-far-overlooked surface-assisted orbital ordering may prevail in transition metal oxides, heavy fermion superconductors and other materials.

MAIN TEXT

The newfangled orbital-mediated quantum phenomena have proved over the last decade to be far-reaching and more complex than before (*1-3*), exhibiting exotic orbital orders (*4, 5*), nontrivial orbital-fluctuation-mediated superconductivity (*6-9*), and orbital Kondo effect (*10, 11*). To unravel electronic, spin, and orbital correlations in these phenomena, it is crucial to have a direct, real-space, access to orbital texture, but so far orbital-sensitive probes have shown rather limited functionality. Recently, several groups demonstrated the orbital-selectivity of a scanning tunneling microscope (STM) within a fine-tuned tip-sample distance (TSD) (*12-14*). According to the Tersoff-Hamann theory (*15*), an STM image corresponds to a contour mapping of local density of states (DOS) at the center of the tip apex curvature. When the tip is away from the surface (< 1 nm), it probes the electronic states that extend farther from the surface, i.e., valence states composed of the *s* and *p* orbitals of constituent atoms. When the tip is close, it becomes sensitive to inner core orbitals like the *d* orbitals, which fall off at short distances from the surface. A well-controlled calibration of TSD allows us to image individual orbital states (*12-14*). Here, we exploit the orbital sensitivity of STM to unveil a surface-assisted cobalt *d*-orbital order in the heavy fermion compound $CeCoIn_5$. This compound is naturally born at the quantum critical point having the unconventional *d*-wave superconducting phase below $T_C$=2.3 K (*16*). The compound has a tetragonal $HoCoGa_5$ structure with lattice constants a = 0.46 nm and c = 0.76 nm as shown in Fig. 1A. Along *c* axis, three different layers, Co, In, and CeIn layers, stack in the order of Co-In-CeIn-In-Co.

Figures 1B and C show overview images of two typical (001) surfaces of $CeCoIn_5$ cleaved *in-situ*. The narrow terraces on both surfaces are separated by a step of ~ 0.76 nm, which equals to the lattice constant along the *c* axis, indicating the same layers covering the whole surface area. The surface in Fig. 1B has a relatively large number of cone-shaped and pit-shaped defects and adsorbates. In contrast, the surface in Fig. 1C is smooth except occasional line defects, indicated by red arrows, and surface standing waves around the steps. The visible atomically resolved square lattice on the terraces has an interatomic distance of ~ 0.5 nm equal to the lattice constant of the CeIn or Co plane (see also Fig. 1D). Although the CeIn layer has two atom species, identical atomic arrays in the CeIn and Co layers have been observed in previous studies (*17-19*). The two layers were claimed to be distinguishable from the shape of the tunneling spectrum (*17*). However, the viability of such correspondence has been questioned (*20*); in any case, we could not draw any conclusion within our measurements (Fig. S1 of the Supplementary Materials). On the other hand, these two layers can be unambiguously distinguished once we identify the missing atoms in the CeIn layer. According to previous reports on $CeCo(In_{0.9985}Hg_{0.0015})_5$ (Fig. S6 of Ref. 17), the Hg impurities, which can only substitute for In atoms, partially occupy the top sites of the original square lattice in a CeIn plane. This leads us to conclude that the visible atoms within a normal TSD (tunnel resistance > 2 MΩ) are In.

We used the orbital sensitivity of STM to visualize the missing Ce atoms. We took STM images on the two surfaces in Fig. 1 for various tunneling currents $I_T$ and sample bias voltage $V_S$ fixed to 10 mV. Figures 2A and 2B show drastic changes in surface topographies as a function of $I_T$. The white box indicates the unit cell. At $I_T$ =100 nA (Fig. 2A), on the terraces in Fig. 1B appear new bright protrusions in-between the original atomic sites. These new protrusions are most likely Ce atoms of the CeIn plane (images of Ce 5*d* orbitals) because no extra atoms exist in the Co plane. By contrast, no appearance of additional atoms on the terraces in Fig. 1C evidences that these surfaces are Co layers.

On the terraces in Fig. 1C, the round shapes of surface Co atoms gradually transform into elongated dumbbells with $I_T$ increasing from 1 nA to 100 nA as shown in Fig. 2B. The fact that these dumbbells between adjacent atomic sites alternate in the [100] and [010] directions, breaks the original periodic symmetry of Co atoms observed at low $I_T$ (See Fig. S2 and Movie S1 of the Supplementary Materials, for more details of this evolution of Co atom appearances in STM topographies). Such a symmetry breaking can hardly be interpreted as resulting from local tip-sample interactions. At all times our STM was operated in the tunneling regime, since the tunneling current $I_T$ as a function of TSD has an exponential dependence up to ~ 160 nA (Fig. 2C). To ensure further that this ordered structure is not a tip-induced artifact but truly reflects the surface DOS at a close distance, we have reproduced these results by repeating the experiment with three different tips and on two different cleaved surfaces (see the Material and Method and Note S1 of the Supplementary Materials for more details). Tip-induced electric fields are negligible for a bias voltage at 10 mV.

Co 4$s$-electrons are not localized and their wave functions extend to the vacuum, making the round shapes of Co atoms when the tip is far away (Fig. 2B, left panel). When the tip is close, the electron tunneling to inner Co 3$d$ orbitals becomes possible (See Fig. S3 of the Supplementary Materials). Among them, the $d_{xz}$ and $d_{yz}$ orbitals can cause dumbbell shapes in STM topography, as is the case for a similar dumbbell shape observed on a single Co atom (*21*). The two 3$d$ orbitals are degenerate and partially occupied in tetragonal CeCoIn$_5$, their degeneracy being lifted by the on-site Coulomb interactions at surface as will be discussed later. We conclude that these dumbbells are a direct visualization of the Co $d_{xz}$ and $d_{yz}$ orbitals with unbalanced electron occupations alternating at adjacent Co sites and that the staggered dumbbell pattern is due to antiferro-type orbital order (OO) involving the two orbitals.

A detailed investigation of the line defects on the terraces in Fig. 1C provides an evidence indicating that the observed OO occurs only in a surface layer. We compared three line profiles along atomic rows with an identical length, shown in Fig. 3A. One is taken within a domain (c), and the other two are taken across the line defect (a, b). By comparing these three line profiles, we see that the positions of the protrusions have no offset across the defect (Fig. 3b), thus excluding any structural mismatch as an origin of the defect. When the tip is far away from the surface, the protrusions show no offset across the defect (Fig. 3B), which excludes any structural mismatch as an origin of the line defect. By contrast, when the tip is close, the staggered dumbbell pattern, which appears on both sides of the defect (Fig. 3C), has a phase mismatch across the defect (Fig. 3D). This shows that the line defect must result from the domain formation of the ordered structure. These defects once again support the fact that the OO is not a tip-induced artifact, since they are still visible even when the tip is far away. Figure 1C shows that the line defects are not connected across the steps, thus strongly indicating that they are not three-dimensional domain boundaries, and that the OO occurs within a single Co plane. Since no signature of such $d$-OO has been observed so far with surface-insensitive bulk methods (*16, 22-25*), the OO is most likely confined in the vicinity of the surface.

To gain more insight, we have performed first-principles calculations within the generalized gradient approximation (GGA, GGA+$U$) using Wien2k code (*26-28,* Materials and Methods). Figures 4A and 4B show the DOS of Co1 atom in the bulk and on the surface of CeCoIn$_5$, obtained respectively from bulk (GGA) and slab (GGA+$U$ with $U_{Co}$=4 eV and $U_{Ce}$=5 eV) calculations (*29*). The positive and negative sides of the DOS represent the majority and minority spin contributions, respectively. The $d_{xz}$ and $d_{yz}$ orbitals remain degenerate in the bulk without noticeable spin polarization (0.11 $\mu$B) in the $d$ orbitals (Fig. 4A). Still, they are antiferromagnetically coupled to each other and ferromagnetically coupled to the Ce spins. In the following slab calculations, we focus on this magnetic pattern. Interestingly, the degeneracy of the $d_{xz}$ and $d_{yz}$ orbitals is lifted with almost fully occupied majority-spin orbitals and partially empty minority-spin orbitals (Fig. 4B), ending with a finite spin polarization (2.00 $\mu$B) on the surface. In particular, the $d_{xz}$ orbital is almost fully occupied while the $d_{yz}$ orbital is half-filled. In the neighboring Co2 atom, we obtained opposite electron occupation in the two orbitals; so ensures a staggered $d_{xz}$-$d_{yz}$ OO pattern. Figures C, D, and E illustrate the anisotropy in the charge density around the Fermi energy at a Co terminating surface. The charge density is plotted in xy planes located at 0.10 nm (D) and 0.25 nm (E) above the terminating Co plane. In the in-plane charge density, the OO pattern is only visible at a short distance to the surface (Fig. 4D), in agreement with the TSD-dependent STM images in Fig. 2B.

Although other magnetic patterns of Co spins can compete with the one considered here, the OO, which can facilitate virtual hopping processes to gain energy, is found to be more stable than the phase without OO but having the same magnetic order. Other effects of smaller energy scale, such as the Kondo effect, could alter to some degrees the DOS near the Fermi energy. However, the unbalanced occupation of $d_{xz/yz}$ orbitals persists in the entire energy window, of the order of several eV (see Fig. 4B), so the OO feature cannot be easily suppressed.

The mechanism of the OO observed experimentally and obtained numerically can be explained as follows. The number of conduction electrons at surfaces is generally reduced significantly owing to the reduction of coordination numbers. This leads to a considerable reduction of electrostatic screening. As a result, the Coulomb repulsion between conduction electrons increases (*30*). With on-site Coulomb repulsion enhanced, a single Co atom in a Co-terminating surface of CeCoIn$_5$ can develop a Mott gap with a tendency to lift the degeneracy between the $d_{xz/yz}$ orbitals. Meanwhile, inter-site Coulomb repulsion provides a driving force to select the OO pattern: the occupied orbitals between adjacent lattice sites tend to avoid aligning with each other in order to lower the Coulomb energy. Consequently, the orbitals are occupied alternatingly into a staggered pattern.

Such a surface-assisted OO observed just now might not be exclusive in the well-known heavy fermion compound CeCoIn$_5$, since orbital degeneracy breaking may occur at any other surfaces for the same reason, although it has escaped detection by current surface techniques. The use of STM to sense the orbitals can open a new era for understanding many-body orbital correlations and unveiling order parameters of poorly understood phenomena such as a hidden order in URu$_2$Si$_2$ (*31*).

# Materials and Methods

## Experiments

High-quality single crystals of CeCoIn$_5$ were grown by an indium self-flux method as described in Ref. 15. In this study, we used five single crystals. The samples qualities were tested by magnetization measurements using a Quantum design SQUID magnetometer down to 2.0 K prior to STM measurements. STM/STS measurements were performed with a $^3$He cryostat-based ultrahigh vacuum STM (Unisoku, USM-1300S) and a STM controller (Specs, Nanonis). Magnetic fields were applied perpendicular to the cleaved surface.

Atomically flat surfaces parallel to the *ab*-plane were obtained by cleaving at room temperatures under ultrahigh vacuum condition of ~ 10$^{-8}$ Pa. We reported results on a total of nine cleaves. After approaching a cleaved surface, we moved around by changing a full scan area with the XY moving stage to look for a flat clean area (successful rate is 44%). On two out of the four cleaved surfaces, we could find only CeIn plane-dominating areas. On the other two surfaces, we found both CeIn and Co planes. We also observed a pure Indium area stuck to surface steps. We performed STM measurements on all successfully cleaved areas by using three mechanically sharpened PtIr tips. All of the tips reproduced clear images of the Ce atoms on CeIn planes and/or the evolution of dumbbell structures of Co atoms on Co planes.

To check the stability of the dumbbell structure against temperature and external magnetic fields, we performed STM measurements on a Co plane with TSD in an external magnetic field up to 5.5 T (> $H_{C2}$) and at temperature from 500 mK to 6 K (> $T_C$). We obtained identical structures within our noise level for all measured field and temperature ranges.

## First principles calculations

The first-principles calculations were performed by Wien2k code based on the all electron, full potential, linearized augmented plane wave method. In the calculations, experimental lattice parameters were used (*16*). The radii (R) of Ce, Co, and In atoms were set to 2.5, 2.5, and 2.39 Bohr, respectively. The plane-wave expansion parameter was chosen as $RK_{max} = 7$. An $8\times8\times7$ mesh of *k*-point sampling was used for the Brillouin zone of the bulk containing two Ce, two Co, and ten In atoms. A $6\times6\times1$ mesh was used for the slab calculation containing eight Ce atoms and ten Co atoms for the five-Co-layer slab. A 10.7-Ångstrom vacuum layer was included in the calculation. The effective $U_{eff}=U-J$ with $J=0$ eV was used for both Ce and Co atoms in the GGA+*U* calculations. The adopted supercell allows nearest neighboring Ce-Ce atoms to couple antiferromagnetically, in consistency with experiments. We have performed the calculations with and without spin-orbit coupling and presented the results without spin-orbit coupling since the same conclusion can be reached.

**Supplementary Materials**

Fig. S1. Tunneling spectra on cleaved $CeCoIn_5$ surfaces.

Fig. S2. Current-dependent cross-sectional profiles of Ce and In atoms on a Ce- In plane (a), and those of Co atoms on a Co plane (b) with corresponding topographic images.

Fig. S3. Charge density profiles of the Co atom orbitals on the Co termination.

Movie S1. Evolution of dumbbell structure by decreasing the tip-sample distance.

Note S1. Excluding the possibility of tip-induced artifacts in dumbbell formation

Note S2. Topographic similarity between our dumbbell ordered structure and the images in Ref. 14 and their difference in underlying physics

**Acknowledgments**

We thank M. -T. Suzuki and N. Tateiwa for fruitful discussions and collaborations in the initial stage of the work. We also acknowledge discussions with S.-Y. Shiau, R. Peters, S. Kim, T. Miyamachi, F. Komori, M. Haze, S. Yamamoto, T. Kawae, K. Fujita, C. K. Kim, E. Minamitani, Y. Okada, H. Sakai, and A. Miyake. **Funding:** This work is partially supported by Grants-in-Aid for Scientific Research from the Japan Society for the Promotion of Science (Nos. 25707025, 26110507, 26120508, and 16K17744). H.L. acknowledges the Singapore National Research Foundation for the support under NRF Award No. NRF-NRFF2013-03. **Author contributions:** Y.Y. conceived and directed the project. H.K. designed experiments. H. K. and Y.Y. carried out the experiments and analyzed the data. Y. Haga, Z.F. synthesized and characterized the single crystals. C.C.L., T.R.C., H.T.J., H.L. performed theoretical calculations. Y.Y., C.C.L., and H. L. wrote the manuscript. Y. Hasegawa supervised the project. All authors commented to the manuscript. **Competing interests:** The authors declare that they have no competing interests.

**Figures and Tables**

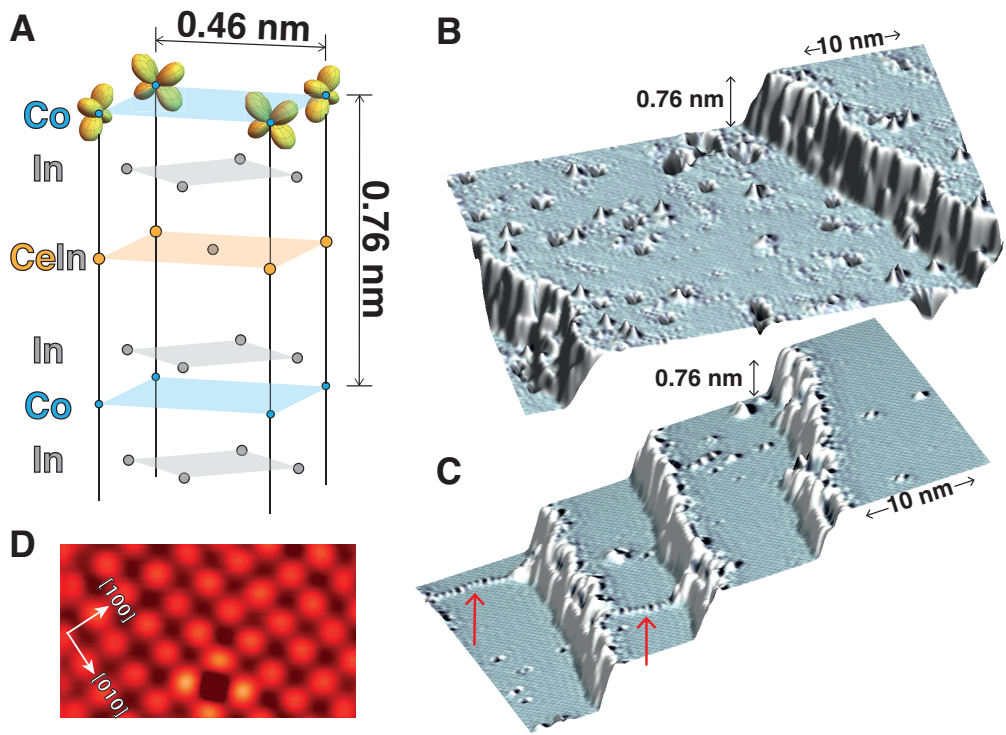

**Figure 1 | The crystal structure and topography of cleaved surfaces of CeCoIn$_5$. (A)** Schematic of the tetragonal crystal structure of CeCoIn$_5$ with $d_{xz}$-$d_{yz}$ orbital orders at the topmost cobalt plane. **(B, C)** Overviews of two-typical cleaved surfaces in this study. Narrow terraces (10~50 nm) are separated by a step of ~ 0.76 nm. The topographic images are colorized with their derivatives to emphasize the atomically resolved lattice structures. Red arrows in (C) indicate line defects (A: tunneling current $I_T$=50 pA, sample bias voltage $V_S$=50 mV, B: $I_T$=15 pA, $V_S$=10 mV). **(D)** Typical atomically resolved STM image taken on the surface of (B) ($I_T$=1 nA, $V_S$=50 mV). We observed an almost identical atomic-lattice image on the surface of (C), as will be shown in Fig. 2B.

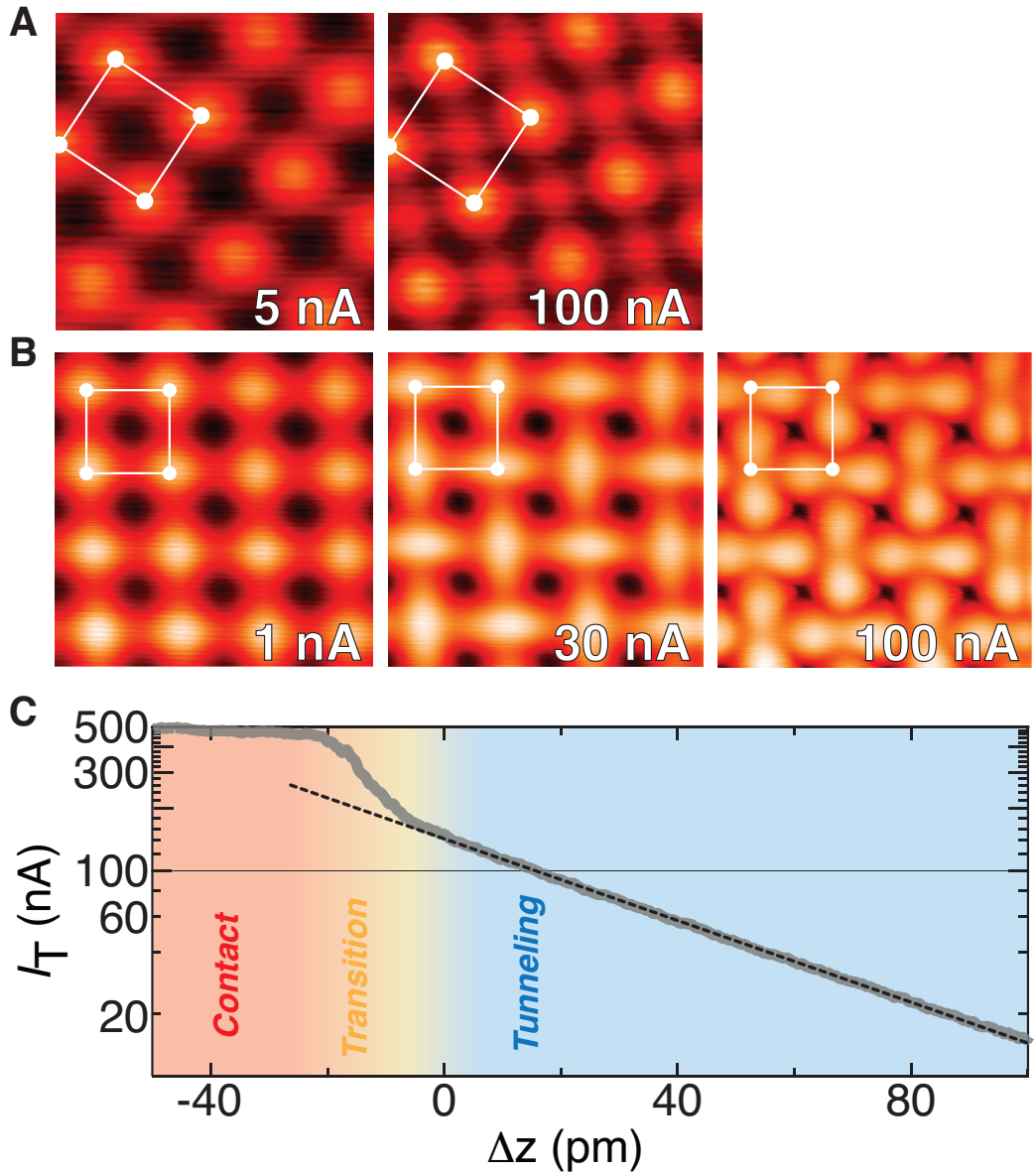

**Figure 2 | Current dependent topographies of the terraces in Figs. 1(B) and 1(C).** The white box in the images indicates the unit cell. **(A)** Atomically resolved topographies of the terraces from Fig. 1B ($V_S$=10 mV, $I_T$=5 nA and 100 nA at $T$= 500 mK). With increasing the current set point, new atoms appear in-between original atomic sites with an interval of ~ 0.5 nm. **(B)** Atomically resolved topographies of terraces in Fig. 1C ($V_S$=10 mV, $I_T$=1 nA, 30 nA, and 100 nA at $T$=1.7 K and at $B$=5 T). An external out-of-plane magnetic field is applied merely to have better stability for STM measurement. With $I_T \gtrsim 10$ nA, the shape of atoms gradually changes to a dumbbell with two lobes. The angle between the dumbbells at adjacent sites is 90º. **(C)** Current trace as a function of tip approaching distance $\Delta z$, measured at an atomic site in (B) with $V_S$=10 mV. $I_T$ shows an exponential dependence up to ~ 160 nA, indicating the tunneling regime.

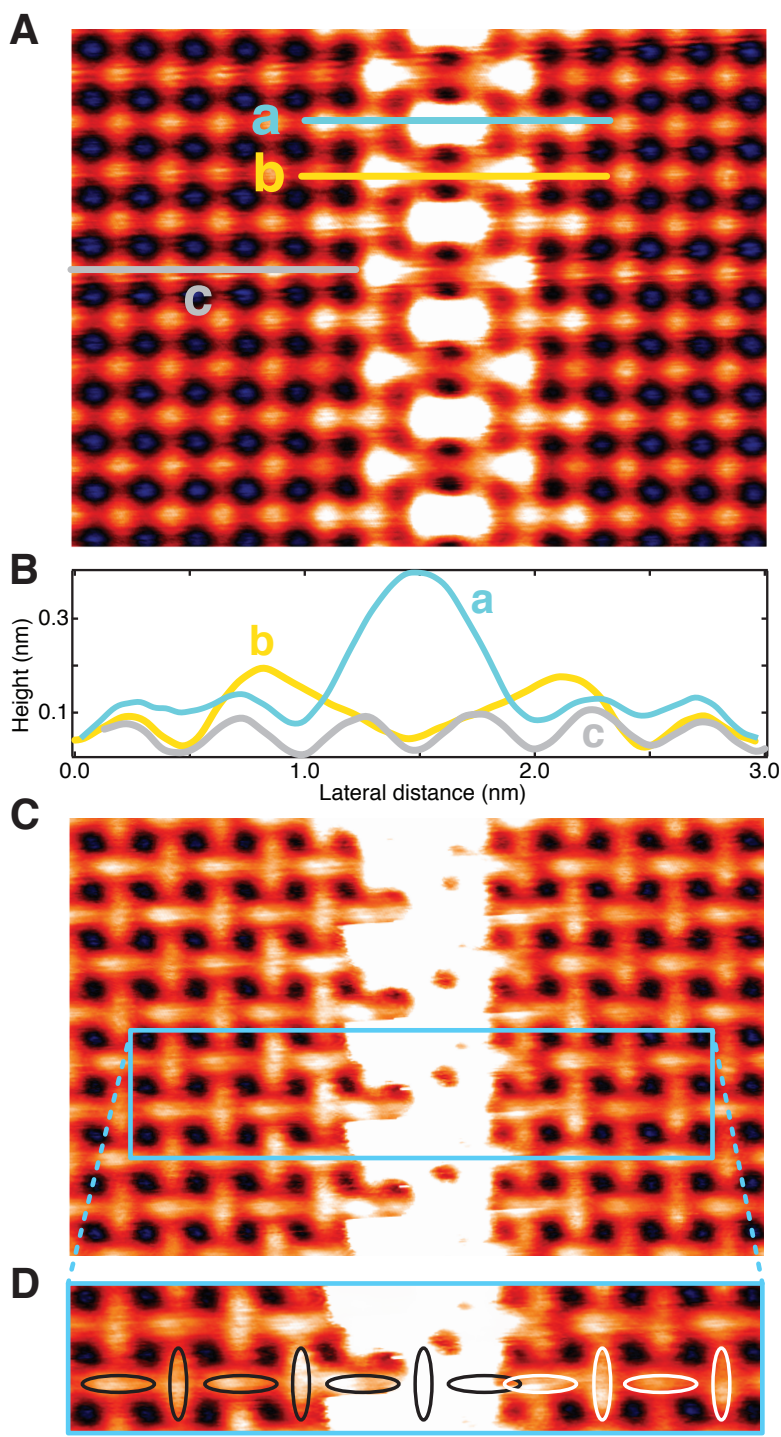

**Figure 3 | Current-dependent topographies of the surface in Fig. 1(C) with a line defect.** **(A)** Topography of the terrace at 1.7 K with a line defect along the vertical axis of the image ($I_T$=10 nA, $V_S$=10 mV). **(B)** Line profiles indicated as lines a, b, and c in (**A**). Atomic positions in profiles a and b matches very well with the ones in profile c in the both sides of the line defect, indicating that no lattice distortion is present across the line defect. **(C)** Topography in the same field of view with (A) but with a shorter tip-sample distance ($I_T$=100 nA, $V_S$=10

mV). Dumbbell shapes are now visible. **(D)** Magnified image of the rectangle area indicated with a sky-blue rectangle in **(C)**. Ellipses help visualize the change in the dumbbell arrangement across the line defect.

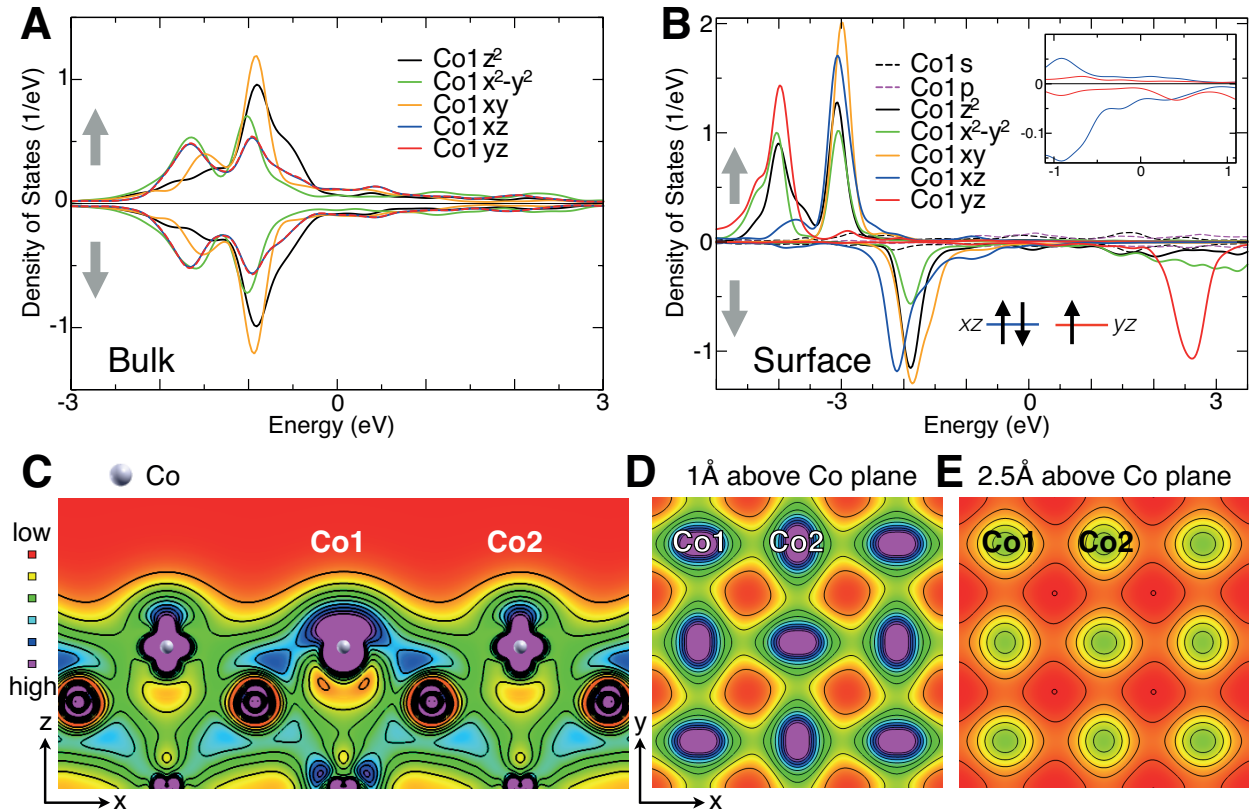

**Figure 4 | First-principles electronic structures. (A, B)** Density of states of *d* orbitals of bulk (A) and surface (B) Co atoms of $CeCoIn_5$. Positive (negative) values show the majority- (minority-) spin contribution. The inset of (B) shows the densities of states of $d_{xz}$ and $d_{yz}$ orbitals around the Fermi energy, demonstrating unbalanced orbital occupations. **(C-E)** The integrated charge density at the slab surface between -0.2 and 0.0 eV in the xz and xy planes. The xy planes are chosen at 0.10 nm (D) and 0.25 nm (E) above the top-most Co plane. The maximum value to present the color-coded plot is chosen as 10, 5, and 1 $e/nm^3$ for (C), (D), and (E), respectively.

# Supplementary Materials for

## Atomic-scale visualization of surface-assisted orbital order

Howon Kim, Yasuo Yoshida*, Chi-Cheng Lee, Tay-Rong Chang,
Horng-Tay Jeng, Hsin Lin, Yoshinori Haga, Zachary Fisk, Yukio Hasegawa

*Correspondence to: yyoshida@issp.u-tokyo.ac.jp

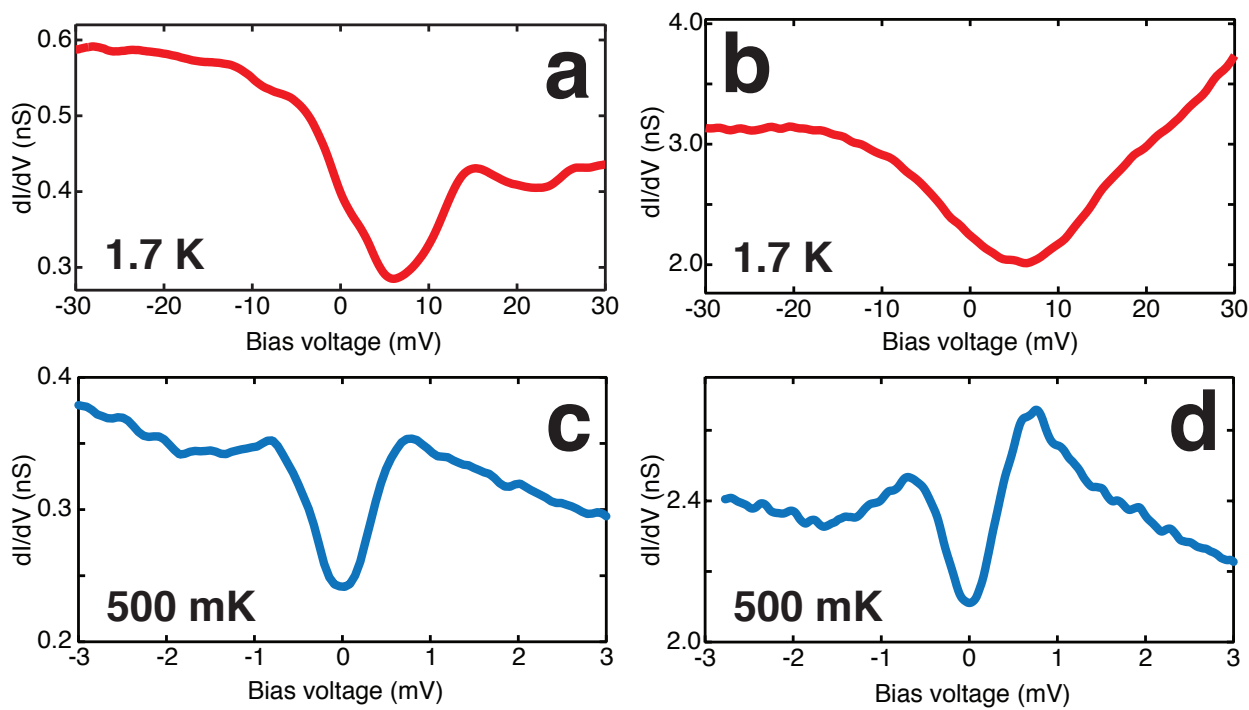

**Figure S1| Tunneling spectra on cleaved surfaces in Figs. 1.(B, C). (a,b)** Both spectra show a gap with a dip around 5 mV as in the case of CeIn planes reported in Ref. 17-19. The difference in shape is probably due to tip status but the overall feature is similar. **(c, d)** Superconducting gap spectrum around $E_F$ taken on the surfaces shown in Figs. 1(**B**, **C**). An identical superconducting gap with $2\Delta \sim 1$ mV is observed on both surfaces with no noticeable difference. Although the CeIn and Co surfaces were claimed in Refs. 17 and 18 to be distinguishable based on the spectral features, we do not find distinct difference in their spectra.

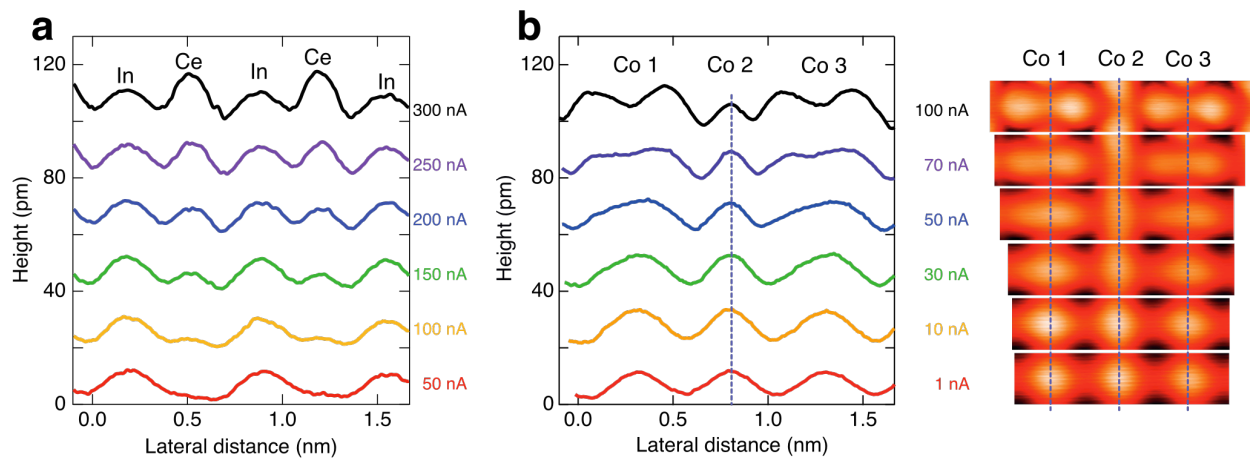

**Figure S2 | Current-dependent cross-sectional profiles of Ce and In atoms on a Ce-In plane (a), and those of Co atoms on a Co plane (b) with corresponding topographic images. a,** Profiles at more than 100 nA are shifted upward for clarity. Enhancement of the Ce height is visible as the current set point is increased. The Ce height becomes even higher than those of In at 250 nA and 300 nA. **b,** Profiles at more than 10 nA are shifted upward for clarity. By increasing the current set point, Co atoms referred to as Co1 and Co3 elongate into two lobes while Co2 shrinks because of the elongation along the perpendicular direction.

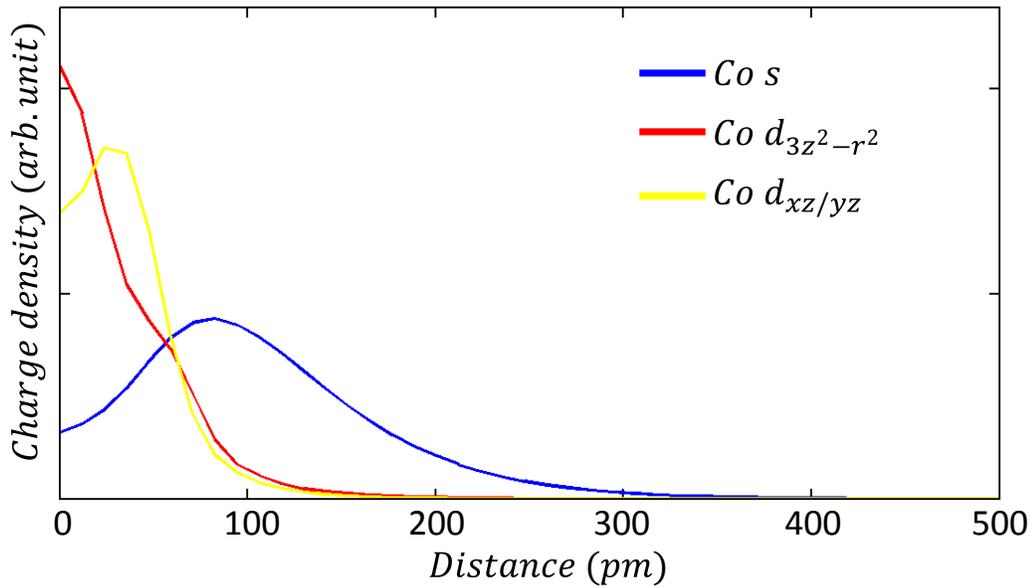

**Figure S3 | Charge profiles of the Co atom orbitals on the Co termination.** The distance on the horizontal axis corresponds to the out-of-the-plane direction. They are obtained from slab calculations, through projection on each spherical harmonics $Y_{lm}$ centered at the atom at hand. The orbital wave functions are extended to a few hundred pm. When the STM tip approaches the Co plane, the $d$ orbitals show up only when the STM tip is very close to the Co atom.

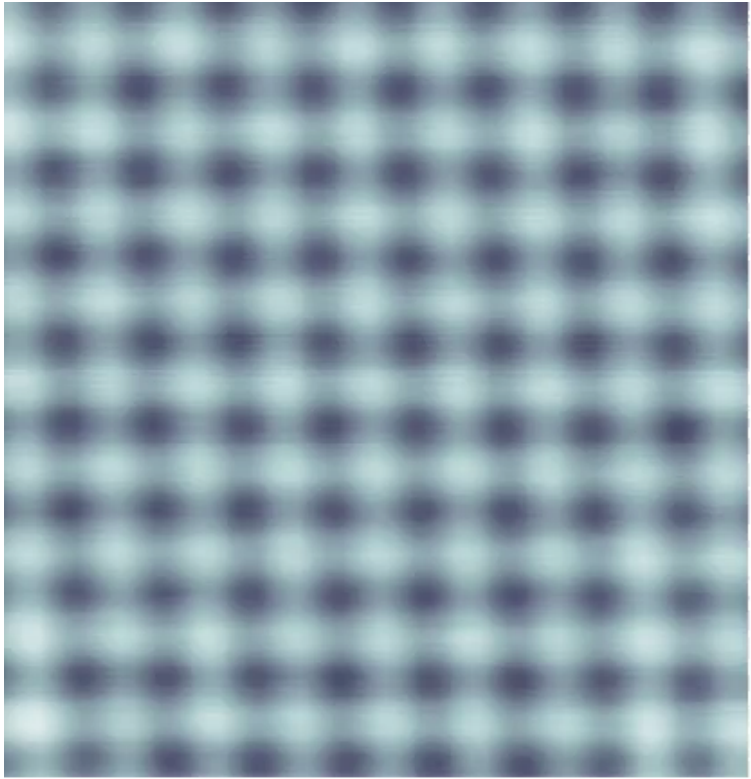

**Movie S1 | Evolution of dumbbell structure by decreasing the tip-sample distance.** The movie demonstrates evolution of the dumbbell ordered structure depending on tip-sample distance with the tunnel resistance between 10 and 91 kΩ.

**Supplemental Note 1: Excluding the possibility of tip-induced artifacts in dumbbell formation**

We have excluded possible experimental artifacts that may affect our observations. As described in the main text, we measured $I_T$ as a function of TSD to ensure that the observations were performed in the tunneling regime. We observed the characteristic exponential dependence of the current up to $I_T \sim 160$ nA on the Co surface (Fig. 2C). We reproduced identical results by repeating the experiment with three different tips and on two different cleaved surfaces. Electric fields caused by small TSDs are negligible because $V_S$ is only 10 mV. These experimental facts rule out the possibilities of any artifacts caused by strong interactions between the tip and sample. Moreover, a detailed investigation of the line defects on the terraces also supports that the dumbbells are not tip-induced artifacts. Since the defects are visible even when the tip is far away from the surface (Fig. 1C), the ordered structure is clearly not a result of the tip-induced artifact.

**Supplemental Note 2: Topographic similarity between our dumbbells ordered structure and the images in Ref. 14 and their difference in underlying physics**

Accessibility to orbital selective tunneling was recently reported on several systems with changing tip-sample distance (TSD) in STM (*12-14*), as mentioned in the abstract. Most recently, Takahashi *et al.*(*14*) reported a topographic evolution from a dot array to an alternating dumbbell array with decreasing TSD on a monoatomic layer of $Fe_4N$ on Cu (001), which resembles the structure we observed on a Co plane of $CeCoIn_5$. However, the physical origins behind the two structures are totally different, although their STM images and tendency in TSD are similar at a glance. In Ref. 14, the alternating dumbbell structure originates from dimerized $d_{z2}$ orbitals of Fe atoms, which are hybridized through the orbitals of N atom. While in $CeCoIn_5$, it stems from the alternatingly selected $d_{xz}$ ($d_{yz}$) and $d_{yz}$ ($d_{xz}$) orbitals of Co atoms on the Co plane, which is due to the lifted degeneracy by on-site coulomb potential enhanced at a surface.